\documentclass{article}

\usepackage{PRIMEarxiv}

\usepackage[utf8]{inputenc} 
\usepackage[T1]{fontenc}    
\usepackage{hyperref}       
\usepackage{url}            
\usepackage{booktabs}       
\usepackage{amsfonts}       
\usepackage{nicefrac}       
\usepackage{microtype}      
\usepackage{lipsum}
\usepackage{fancyhdr}       
\usepackage{graphicx}       
\graphicspath{{media/}}     
\usepackage{siunitx}
\usepackage{float}
\usepackage{bm}
\usepackage{amsmath}
\usepackage{amssymb}
\usepackage{longtable}
\usepackage{caption}

\newcommand{\mbf}{\bm}

\def\R{{ \mathbb{R} }}
\def\N{{ \mathbb{N} }}

\usepackage[colorinlistoftodos, textwidth=65mm, shadow, loadshadowlibrary]{todonotes}

\title{Single CASANOVA? Not in multiple comparisons

}

\author{
  Ina Dormuth$^{1*}$, Carolin Herrmann$^{2}$,
 Frank Konietschke$^{3}$, Markus Pauly$^{1,4}$, Matthias Wirth$^{5,6}$, and Marc Ditzhaus\textsuperscript{\textdagger}$^{7}$\\
$^{1}$Department of Statistics, TU Dortmund University, Dortmund, 
    North Rhine-Westphalia, 44227, Germany \\
$^{2}$Mathematical Institute, Heinrich Heine University Düsseldorf, Düsseldorf, Germany\\
$^{3}$Institute of Biometry and Clinical Epidemiology, Charité – Universitätsmedizin Berlin, 10117 Berlin, Germany \\
$^{4}$Research Center Trustworthy Data Science and Security, UA Ruhr, 44227 Dortmund\\
$^{5}$ Department of Hematology, Oncology and Cancer Immunology, Charité - Universitätsmedizin Berlin, Berlin, Germany\\
$^{6}$ Department of General, Visceral and Pediatric Surgery, University Medical Center Göttingen, Göttingen, Germany.\\
$^{7}$Department of Mathematics,  Otto von Guericke University Magdeburg, Magdeburg, Germany\\
}

\begin{document}
\maketitle

\begin{abstract}
When comparing multiple groups in clinical trials, we are not only interested in whether there is a difference between any groups but rather the location. Such research questions lead to testing multiple individual hypotheses. To control the familywise error rate (FWER), we must apply some corrections or introduce tests that control the FWER by design. In the case of time-to-event data, a Bonferroni-corrected log-rank test is commonly used. This approach has two significant drawbacks: (i) it loses power when the proportional hazards assumption is violated  \cite{bardo2023} and (ii) the correction generally leads to a lower power,
especially when the test statistics are not independent \cite{konietschke2013multiple}. 
We propose two new tests based on combined weighted log-rank tests. One as a simple multiple contrast tests of weighted log-rank tests and one as an extension of the so-called CASANOVA test \cite{ditzhausCASANOVAPermutationInference2021}. The latter was introduced for factorial designs. 
We propose a new multiple contrast test based on the CASANOVA approach. Our test promises to be more powerful under crossing hazards and eliminates the need for additional p-value correction. We assess the performance of our tests through extensive Monte Carlo simulation studies covering both proportional and non-proportional hazard scenarios. Finally, we apply the new and reference methods to a real-world data example. The new approaches control the FWER and show reasonable power in all scenarios. They outperform the adjusted approaches in some non-proportional settings in terms of power. \end{abstract}


\section{Introduction}

Time-to-event or survival analysis is essential across medical research, engineering, and social sciences. Trials often involve multiple groups (treatment arms) or factorial designs, creating unique statistical challenges. The primary research focuses not merely on whether any arms differ but specifically identifying which groups show differences. Thus, traditional global test procedures like ANOVA-type methods, which test null hypotheses of equal hazard ratios or cumulative hazard rate functions, are often inadequate \cite{Konietschke2012, ditzhausCASANOVAPermutationInference2021}. Instead, flexible multiple comparison procedures are crucial in modern data analysis. Current approaches typically employ pairwise multiple log-rank tests with adjustments for multiplicity (e.g., Bonferroni correction) \cite{logan2005pairwise}, but these methods can lack efficiency due to restrictive assumptions about the correlation structure of test statistics \cite{Gao2008,gao2008nonparametric}. In recent years, many researchers developed \textit{multiple contrast test procedures} (MCTPs) along with simultaneous confidence intervals (SCIs) (usually conducted as maximum tests), which are valid for arbitrary correlations of the test statistics and use the correlation within the multiplicity adjustment for various endpoints (means, proportions, Mann-Whitney effects) \cite{Bretz2001,schaarschmidt2009asymptotic,hasler2008multiple,konietschke2013multiple,blanche2022closed}. Munko et al. \cite{munko2024} introduced a restricted mean survival time (RMST)-based multiple contrast tests for time-to-event data. Since the RMST should not be employed under crossing hazards \cite{dormuth2022test, dormuth2023comparative}, we aim to close this gap and introduce a powerful and flexible MCTP for analyzing survival data with crossing hazards.

The log-rank test is one of the most prominent test procedures in survival analysis. The method is well known to be optimal when the proportional hazards (PH) assumption is met but significantly loses power otherwise \cite{dormuth2023comparative}. Even though the problem is fairly well known, a substantial amount of investigators of (clinical) trials still ignore the issue and publish their findings upon log-rank tests in leading high-quality peer-review journals even when the assumption is violated \cite{kristiansen2012prm39,trinquart2016comparison, dormuth2023comparative}. For the analysis of two independent samples, weighted log-rank tests and their combinations comprise a great alternative to the classical log-rank test and are beneficial in non-proportional hazards models \cite{andersen2012statistical, fleming2013counting, BrendelETAL2014, DitzhausFriedrich2018}. Ditzhaus and Friedrich \cite{DitzhausFriedrich2018} propose a Wald-type test of multiple weight functions within a single multivariate test. Which weight function to choose depends on the alternative of interest and cannot be recommended in a general way. However, the test does not provide information on which weight function appears most powerful. For the analysis of more than two samples and factorial designs, Ditzhaus et al. \cite{ditzhausCASANOVAPermutationInference2021} extended these procedures to the \textit{Cumulative Aalen Survival Analysis-of-Variance} (CASANOVA) method. In principle, they are global ANOVA-based tests (quadratic forms) and can be used to estimate and test main and interaction effects in general factorial designs. Estimating and testing user-specific contrasts are impossible, limiting their application in statistical practice. To overcome these shortcomings, we propose a novel flexible MCTP. Extensive simulation studies indicate that the test is more powerful under non-proportional hazards and eliminates the need for additional p-value correction. The remainder of the paper is organized as follows. The second section introduces the main statistical methods employed in the analyses. The third section describes the simulation setup and the corresponding results. The following section applies the methods of interest to a real-world data example. The final conclusions are drawn in section five, together with future research questions.

\section{Set up}\label{sec:setup}

Multiple contrasts are faced in many research questions related to time-to-event endpoints. Applying separate tests without adjusting for multiple testing increases the likelihood of false discoveries and inflated error rates. In the following, we present different well-established statistical methods for an underlying multiple contrast problem with time-to-event endpoints, as well as our newly developed method based on a combination of multi-directional log-rank tests and the concept of maximum tests.

\subsection*{Statistical model }
First, we define the underlying statistical model. Therefore, we consider a study design involving $k\geq 2$ groups (treatment arms) of $n_j$ independent subjects, each with time-to-event data $T_{ji}$ and right-censoring time $C_{ji}$. The statistical model considered here can be summarized by mutually independent positive random variables  $T_{ji}\sim F_j,\quad \text{and} \quad C_{ji}\sim G_{j}, \quad j=1,\ldots,k;~ i=1,\ldots,n_j,
$
where $F_j$ and $G_j$ are both continuous distribution functions, respectively. Furthermore, let  $X_{ji}=\min(T_{ji},C_{ji})$ denote the observed time and $\delta_{ji}=I(X_{ji}=T_{ji})$ the censoring status with $I(\cdot)$ being  the indicator function. The statistical model considered here does not entail any parameters but rather the survival distributions that could be used to define reasonable treatment effects. The cumulative hazard rate function for group $j$ is defined by
\begin{eqnarray}
  A_j(t) = \int_0^t (1-F_j(x))^{-1}dF_j(x), t\geq 0, \, j=1,\ldots,k.
\end{eqnarray}

We further assume non-zero sized groups by $n_j/n\to \kappa_j\in(0,1)$ as $\min(n_j:j=1,\ldots,k)\to \infty$ and we exclude the case of only censored values within one group by assuming that $0 < F_j(t) < 1$ and $0 < G_j(t) < 1$  $\forall j= 1,...,k$ and some $t>0$. 

\subsection*{Multiple null hypotheses}
The cumulative hazard rate function of treatment arm $j$ called $A_j(t)$, summarizes the total accumulated risk of experiencing the event that has been gained by progressing to time $t$. No difference (i.e., no effect) between treatment arms $j_1$ and $j_2$ with $j_1 \neq j_2$, corresponds to $A_{j_1}(t)\equiv A_{j_2}(t)$, or, equivalently, $A_{j_1}(t) - A_{j_2}(t)\equiv 0$. In the several sample problems, let $ \mbf{H} \in \mathbb{R}^{q\times k}$
be a contrast matrix satisfying $\mbf{H}\mbf{1_k} = \mbf{0}_q$ with $\mbf{1_k}$ and $\mbf{0}_q$ denoting vectors of ones and zeros, respectively. We denote the entries of $\mbf H$ as $h_{j_1, j_2}$. For ease of presentation, we focus on the two-sample problem. Here, the most prominent matrices are the ones of Dunnett- and Tukey-type. The entries are composed of a single $-1$ and $1$, indicating the two sample comparisons of interest. We define the corresponding index sets $I_{\text{Dunnett}} = \{(1,2), \ldots, (1,k) \}$ and $I_{\text{Tukey}} = \{(1,2), \ldots, (1,k), (2,3), \ldots, (2,k), \ldots, (k-1,k) \}$. In the following, we will indicate the position in the matrix or vector by the corresponding indices $j_1$ and $j_2$, for example, $(-1, 1, 0, \ldots, 0) = \mbf{h}_{1,2}$ for $j_1=1$ and $j_2=2$.

The hypotheses we seek to infer are expressed in relation to the cumulative hazard rate functions as follows:

\begin{align*}
\label{eqn:null}
\mathcal H_0 &: \mbf{H} \mbf{A} = \mbf{0}_q, \quad\mbf{A}=(A_{1},\ldots,A_k)^{\top},\\
H_0^{j_1j_2} &:  \{\mbf{h}_{j_1j_2} \mbf{A} = 0\}, ~(j_1,j_2) \in I,
\end{align*}

with $\mbf{A}^{\top}$ denoting the transposed vector of $\mbf{A}$ and 
$I$ being either 
$I_{\text{Dunnett}}$ or $I_{\text{Tukey}}$.
In general, the contrast matrix selection depends on the specific question of interest underlying the analysis. 

\section{Statistical Tests}
\subsection*{Adjusted log-rank}
As a reference method, we consider the Bonferroni adjusted log-rank test. Therefore, we define the Bonferroni-adjusted significance level $\alpha_{\text{Bonferroni}} = \alpha / q$
where $\alpha$ is the original significance level and $q$ is the number of comparisons. The Bonferroni adjustment for multiple comparisons in a survival setting is a standard procedure in clinical settings, as discussed in Logan et al. (2005) \cite{logan2005pairwise}. The authors have provided a comprehensive description and suggested various methods for adjusting the number of comparisons.

We define the weighted log-rank test as a generalization of the classical log-rank test. Therefore, we employ the conventional counting process notation. Let \( N_j(t)=\sum_{i=1}^{n_j}I\{X_{ji}\leq t, \delta_{ji}=1\} \) represent the cumulative number of observed events within group $j$ up to time $t$. Furthermore, we introduce $Y_j(t)=\sum_{i=1}^{n_j}I\{X_{ji}\geq t\}$, which denotes the number of individuals at risk just before time $t$ in group $j$. These counting processes enable us to define the Nelson-Aalen estimator for $A_j$ as $\widehat A_{j}(t)=\int_{0}^t \frac{I\{Y_j(s)>0\}}{Y_j(s)}\mathrm{ d }N_j(s)$ for $j=1,\ldots,k$ and $t\geq 0$.

Then, the weighted log-rank statistic for testing the local null hypothesis $H^{j_1j_2}_0: \{{\mbf h}_{j_1j_2} {\mbf A} = {\mbf 0} \} = \{A_{j_1} = A_{j_2} \} $ can be defined as \cite{andersen2012statistical}:

\begin{align*} 
T_{j_1,j_2}(w) &= T(w, h_{j_1, j_2}) \\
&= \Big(\frac{n}{n_{j_1} n_{j_2}}\Big)^{1/2} \int_0^\infty w\{ \widehat F_{j_1, j_2}(t-)\}\frac{Y_{j_1}(t)Y_{j_2}(t)}{Y_{j_1}(t)+Y_{j_2}(t)} \mathrm{ d }(h_{j_1,j_2} \widehat{\mbf{A}}(t))\\ 
&= \Big(\frac{n}{n_{j_1} n_{j_2}}\Big)^{1/2} \int_0^\infty w\{ \widehat F_{j_1, j_2}(t-)\}\frac{Y_{j_1}(t)Y_{j_2}(t)}{Y_{j_1}(t)+Y_{j_2}(t)} \Big\{\mathrm{ d }\widehat A_{j_2}(t) - \mathrm{ d }\widehat A_{j_1}(t)\Big\}.
\end{align*}

Here, $\widehat F_{j_1, j_2}(t-)$ represents the left-continuous version of the estimator $\widehat F_{j_1, j_2}$, and $ w$ is a continuous weight function and $\widehat{\mbf{A}} = (\widehat{A}_1, \ldots, \widehat{A}_k)^{\top} $. Fleming and Harrington \cite{fleming2013counting} examined a specific subclass of weights $w$ given by $w(t)=t^r(1-t)^g$ $(r,g\in\N_0)$. For instance, when $r=g=0$, the log-rank test is obtained. We derive the individual p-values for the tests from the $\chi^2$ distribution and compare them to $\alpha_{\text{Bonferroni}}$. To make a global statement, we compare the minimal p-value among all local tests to the adjusted significance level.

For practical implementation, we utilize the R package \texttt{survival} and its function \texttt{survdiff}. \cite{survival-package}  

\subsection*{Adjusted mdir}
 In the context of our specific objectives, we are interested in more robust testing procedures towards multiple alternatives. For two group comparisons, the multi-directional log-rank test has been proposed as a combination procedure of different weighted log-rank tests \cite{BrendelETAL2014, DitzhausFriedrich2018}. The test assumes the equality of survival under the null hypothesis, with the choice of weights determining the alternative hypothesis. We are particularly interested in weights that intersect the $x$-axis, such as $w(t_i) = 1 - 2t_i$ as they are specifically designed to address crossing hazard alternatives. 

By default, the R package \texttt{mdir.logrank} \cite{mdir} implements a combination of the log-rank weight $w^{(1)}\equiv 1$ and this crossing weight. Dormuth et al. \cite{dormuth2023comparative} showed that this default set of weights seems to be robust against multiple alternatives. Nevertheless, if desired, additional weights can be combined to cover more alternative hypotheses. With these two weights, the local test statistic takes a studentized quadratic form:

 \[S_{j_1 j_2}=(T_{j_1j_2}(w^{(1)}), \ldots, T_{j_1j_2}(w^{(m)}))\,\hat{\Sigma}^-_{j_1j_2}\, (T_{j_1j_2}(w^{(1)}), \ldots, T_{j_1j_2}(w^{(m)}))^{\top}.\]

The entries of $\hat{\Sigma}_{j_1 j_2} \in \R^{m \times m}$ are given by

\[(\hat{\Sigma}_{j_1j_2})_{p,s}= \frac{n}{n_{j_1}n_{j_2}} \int_{[0,\infty)}w^{(s)}\{\hat{F}_{j_1j_2}(t-)\}w^{(p)}\{\hat{F}_{j_1j_2}(t-)\}\frac{Y_{j_1}(t)Y_{j_2}(t)}{Y_{j_1}(t)+Y_{j_2}(t)}d\hat{A}_{j_1j_2}(t), ~~~ (p,s=1,...,m) \]

with $\hat{A}_{j_1j_2}$ the pooled Nelson-Aalen estimator of groups $j_1$ and $j_2$. $\hat{\Sigma}^-_{j_1 j_2}$ represents the Moore-Penrose inverse of the empirical covariance matrix of the weighted log-rank tests. For linearly independent weights $w^{(1)},\ldots,w^{(m)}$ fulfilling the assumptions of \cite{ditzhausCASANOVAPermutationInference2021} (continuous and of bounded variation), the test statistic $S_{j_1 j_2}$ can be assumed to be $\chi_m^2$ distributed under the null hypothesis. Ditzhaus and Friedrich \cite{DitzhausFriedrich2018} also proposed a permutation-based approach.

Again, we employ the Bonferroni adjusted significance level $\alpha_{\text{Bonferroni}}$ to compare to the obtained local p-values. Analogously to the adjusted log-rank test procedure, we obtain the global test decision by comparing the smallest p-value to the adjusted significance level.

\subsection*{MultiWeightedLR}

Knowing that maximum tests are a common approach for multiple testing problems \cite{konietschkeAreMultipleContrast2013}, a straightforward extension of the weighted log-rank test is to use the maximum over them and exploit the covariance structure between the different tests. 
We use the same weights as for the adjusted mdir approach without combining them in a quadratic form. Instead, we consider each weighted test individually. After calculating the corresponding covariance matrix, we take the maximum of all weighted test statistics as our global maximum test statistic. Mathematically, we write:

\begin{align*}
    T_{\max} =\max_{r \in \{1, \ldots, m\},
    (j_1,j_2) \in I}
    (T_{j_1,j_2}(w^{(r)}))
\end{align*}

For the local testing problem we focus on $ T^{j_1j_2}_{\max} =\max_{(j_1,j_2) \in I} (T_{j_1,j_2}(w^{(r)}))$. Similar to the proof of Theorem 2 in 
\cite{ditzhausCASANOVAPermutationInference2021}, it can be shown that 
the vector $(T_{j_1,j_2}(w^{(r)}))_{r,j_1,j_2}$ is, under regularity conditions, asymptotic multivariate normally distributed with expected value $\mbf 0_{m ^\cdot q}$ and covariance matrix $\widehat{\Sigma} \in \R ^{m \cdot q \times m \cdot q}$. We thus take the equicoordinate $(1-\alpha)$-quantile \cite{Konietschke2012} of this distribution as critical value to obtain the MultiWeightedLR test in the statistic $T_{\max}$.

\subsection*{multiCASANOVA}
Ditzhaus et al. \cite{ditzhausCASANOVAPermutationInference2021} proposed the CASANOVA (Cumulative Aalen Survival Analysis-of-Variance) approach for general factorial designs with right-censored time-to-event data. The core idea of the method is an extension of weighted log-rank tests to the factorial design setup. Therefore, they expanded the combination approach of weighted log-rank tests (mdir) for the two-sample scenario to the general factorial survival designs implemented in the R package \texttt{GFDsurv} \cite{GFDsurv}. For further information we refer to Ditzhaus et al. \cite{ditzhausCASANOVAPermutationInference2021}

We aim to extend the CASANOVA approach to allow the estimation and testing of user-specific contrasts in a multiple testing framework. Compared to the aforementioned approaches, the main difference is that we consider pooled quantities over all groups, not only the two groups of interest. To this end, we define a local test statistic for contrast $\mbf{h}_{j_1, j_2}$ as

\begin{align*} 
\label{eq:wLR}
\Tilde{T}_{j_1,j_2}(w) &= \Tilde{T}(w, \mbf{h}_{j_1, j_2}) \\
&= \Big(\frac{n}{n_{j_1} n_{j_2}}\Big)^{1/2} \int_0^\infty w\{ \widehat F(t-)\}\frac{Y_{j_1}(t)Y_{j_2}(t)}{Y(t)} \mathrm{ d }(\mbf{h}_{j_1,j_2} \widehat{\mbf{A}}(t))\\ 
&= \Big(\frac{n}{n_{j_1} n_{j_2}}\Big)^{1/2} \int_0^\infty w\{ \widehat F(t-)\}\frac{Y_{j_1}(t)Y_{j_2}(t)}{Y(t)} \Big\{\mathrm{ d }\widehat A_{j_2}(t) - \mathrm{ d }\widehat A_{j_1}(t)\Big\},
\end{align*}

where $\widehat F(t-)$ represents the left-continuous version of the pooled estimator $\widehat F$ and $Y(t)$ is the total number of individuals at risk over all groups. As in the adjusted mdir test, we combine several weights (still for one single contrast) by considering the corresponding quadratic form given by

\begin{align*}
    C_{j_1,j_2} = (\Tilde{T}_{j_1,j_2}(w^{(1)}),..., \Tilde{T}_{j_1,j_2}(w^{(m)}))\widehat{\text{Cov}}_{j_1,j_2}^-\, (\Tilde{T}_{j_1,j_2}(w^{(1)}),..., \Tilde{T}_{j_1,j_2}(w^{(m)}))^{\top},
\end{align*}
where the inner matrix is defined by

\[(\widehat{\text{Cov}}_{j_1j_2})_{p,s}= \frac{n}{n_{j_1}n_{j_2}} \int_{[0,\infty)}w^{(s)}\{\hat{F}(t-)\}w^{(p)}\{\hat{F}(t-)\}\frac{Y_{j_1}(t)Y_{j_2}(t)}{Y(t)}d\hat{A}(t), ~~~ (p,s=1,...,m) \]
and $\widehat{\text{Cov}}_{j_1,j_2}^-$ represents its Moore-Penrose inverse. Similar to the maximum approach within MultiWeightedLR, we now consider the maximum of these Wald-type statistics over all contrasts of interest as the global test statistic
\begin{align*}
    C_{\max} = \max_{(j_1,j_2) \in I}
    (C_{j_1,j_2}).
\end{align*}
Note that we did not take the maximum over the different weights as those are already incorporated within the quadratic forms. 

We use the common wild bootstrap 
approach for counting processes in time-to-event analyses \cite{bluhmki2019wild, bluhmki2018wild} to approximate the limiting distribution. Therefore, we consider independent and identically distributed variables $G_1, \ldots ,G_{n_j}$ with $E(G_i)=0$ and $\text{Var}(G_i)=1$. We obtain the wild  bootstrap version of the Nelson-Aalen estimator by:

\[ \widehat A^*_{j}(t)=\int_{0}^t \frac{I\{Y_j(s)>0\}}{Y_j(s)}\mathrm{ d }\left(\sum_{i=1}^{n_j} G_i N_j(s)\right).\] 

By using $\widehat A^*_{j}(t)$ instead of $\widehat A_{j}(t)$ to derive $\Tilde{T}_{j_1,j_2}$ and thus $C_{j_1,j_2}$ we obtain their wild bootstrap versions $\Tilde{T}^*_{j_1,j_2}$ and $C^*_{j_1,j_2}$, respectively. Since counting processes are discrete, we opt for discrete distributions for the $G_i$. We focus on two common choices: (i) the Rademacher distribution \cite{liu1988bootstrap}, and (ii) the centered Poisson distribution \cite{mammen2012does}. This results in two different wild bootstrap quantiles depending on the distribution of choice: 
$q_{\alpha}^*$ the $\alpha$-quantile of $C_{\max}$ given our data $(X_{ji}, \delta_{ji})$. Then we obtain the global test decision by evaluating $C_{\max} > q^*_{\alpha}$ and the local test decisions by $C_{j_1,j_2} > q^*_{\alpha}$.

\section{Simulation Study}

We conducted an extensive simulation study in R 4.4.0 \cite{R2021} to evaluate the rejection rate and the power performance of the candidate methods. 

\subsection*{Simulation Setup}
We simulated data for $k=4$ groups considering the Tukey- and Dunnett-type contrast matrices. 
We considered four scenarios, each with different distribution functions. Each represents a specific case of hazard relationships such as (i) proportional hazards, (ii) non-proportional and non-crossing hazards, (iii) crossing hazards, and (iv) a mixed scenario. The specific survival functions are presented in Table \ref{tab:Surv}. We set the group size for each scenario to 100; the censoring rates vary between $0\%$ and $30\%$ with uniform censoring. The work of \cite{dormuth2023comparative} indicated that the choice of censoring distribution does not have a major impact on the performance of statistical tests. Considering all possible combinations of censoring, survival distributions, and contrast matrices, we end up with a total of $4$(scenarios)x$1120$ parameter combinations = $4480$ different settings. This is because we only considered the combination of different survival time distributions for the individual groups, but we did not consider the order. This means that $S_1, S_1, S_1, S_2$ is the same combination as $S_1, S_1, S_2, S_1$. For the Tukey-type contrast matrices, we considered every possible comparison for $k=4$ that results in six tests. For the Dunnett-type contrast matrices, we compared the first group to all the other groups, resulting in 3 contrasts.

$10,000$ simulation runs with $1000$ resampling iterations were performed for each setting. The global level of significance was set to $0.05$ throughout. 

\begin{table}[H]
    \centering
        \caption{Simulation scenarios.}
    \begin{tabular}{|c|c c|}
    \hline
    Scenario & CDF & Visualization of the survival and hazard curves \\
    \hline
         Prop &  {$\!\begin{aligned} 
               F_1(t) &= \textit{Exponential}(1.5)  \\   
               F_2(t) &= \textit{Exponential}(2.5)\\
               F_1(t) &= \textit{Exponential}(3.5)  \\   
               F_2(t) &= \textit{Exponential}(4.5) \end{aligned}$} & \raisebox{-.5\height}{
      \includegraphics[width=2.7in]{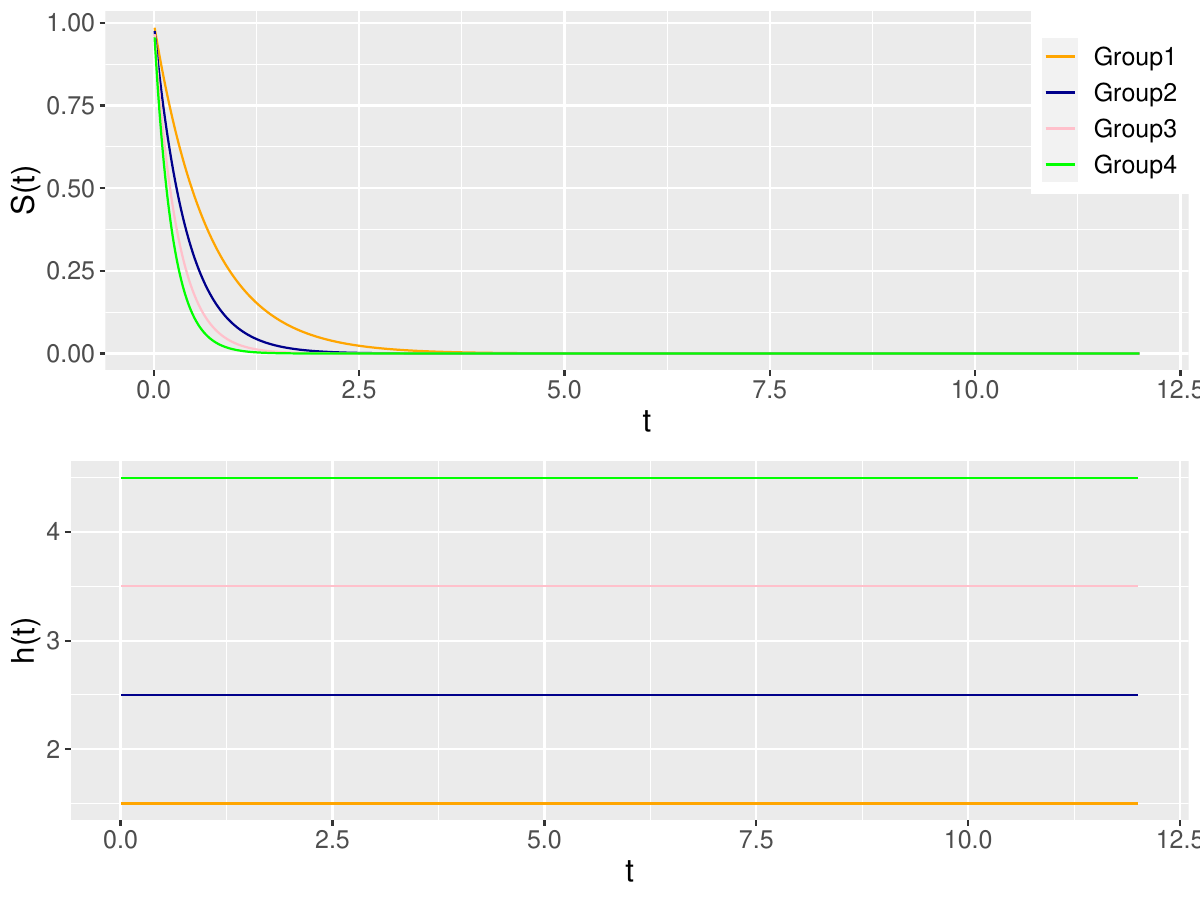}} \\
      \hline
      NProp &  {$\!\begin{aligned} 
               F_1(t) &= \textit{Lognormal}(1.7, 1.7)  \\   
               F_2(t) &= \textit{Lognormal}(2.4, 1.6) \\
               F_3(t) &= \textit{Lognormal}(3.5, 1.7)  \\   
               F_4(t) &= \textit{Lognormal}(4.5, 1.6)  \end{aligned}$} & \raisebox{-.5\height}{
        \includegraphics[width=2.7in]{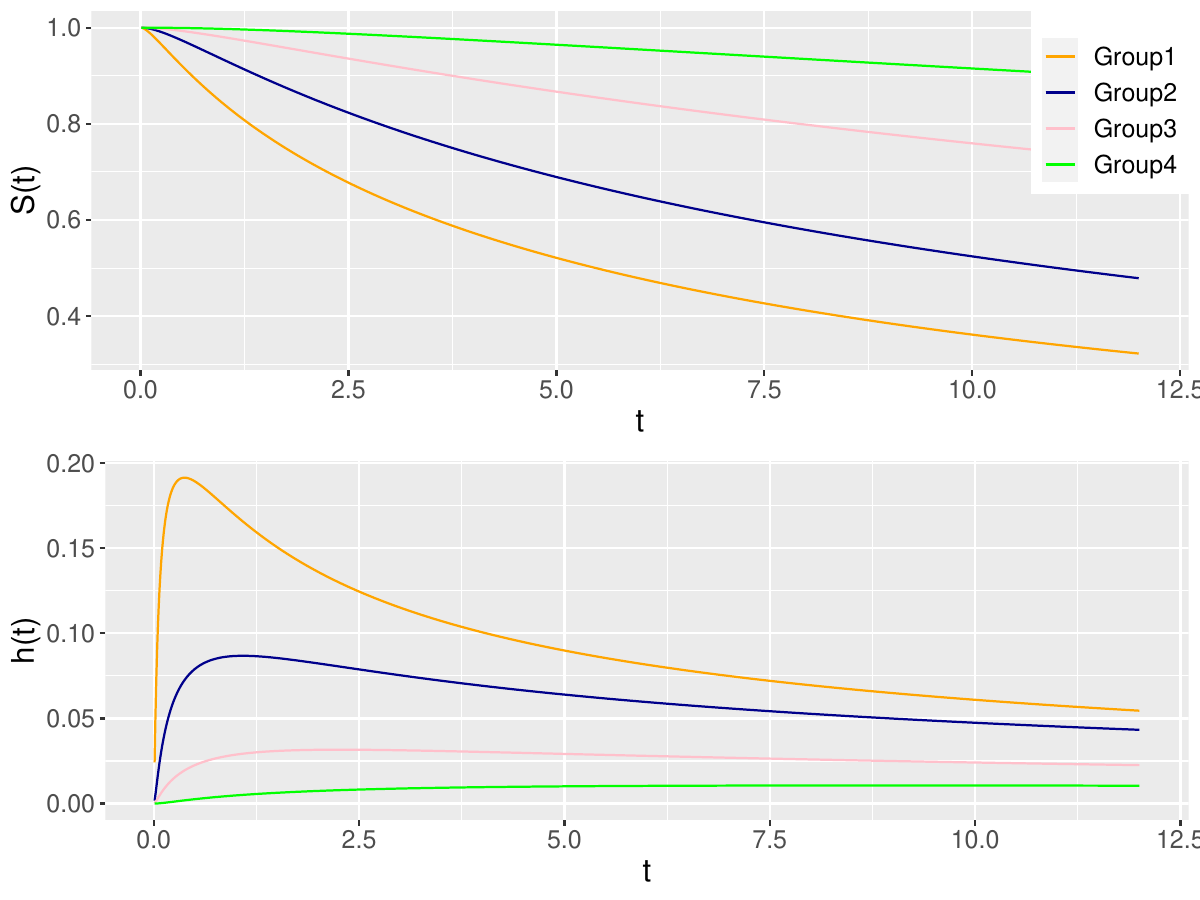}} \\
      \hline
      Cross &  {$\!\begin{aligned} 
               F_1(t) &= \textit{Weibull}(1.5, 5)  \\   
               F_2(t) &= \textit{Weibull}(2.5, 5)  \\
               F_3(t) &= \textit{Weibull}(3.5, 5)  \\   
               F_4(t) &= \textit{Weibull}(4.5, 2.4) 
               \end{aligned}$} & \raisebox{-.5\height}{
      \includegraphics[width=2.7in]{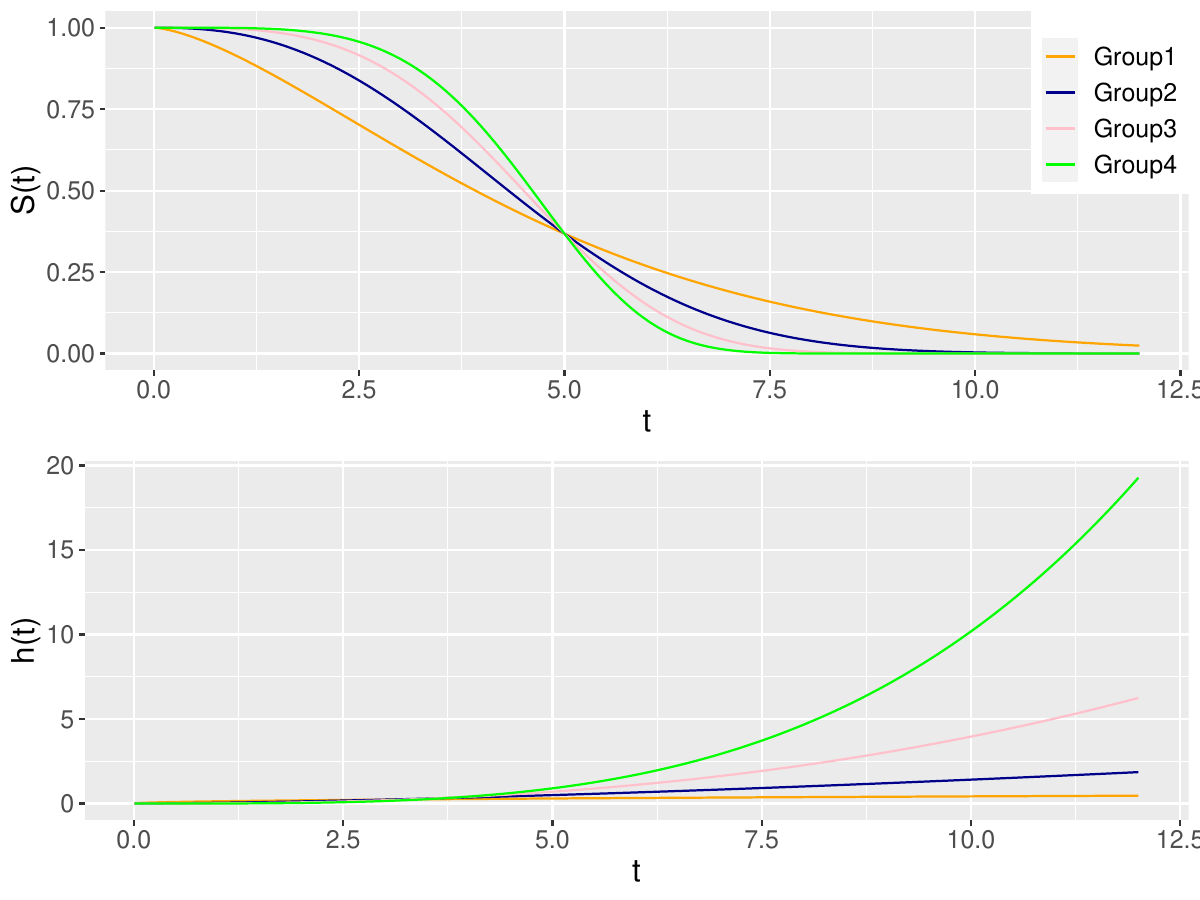}} \\
      \hline
      Mix &  {$\!\begin{aligned} 
               F_1(t) &= \textit{Lognormal}(2.3, 1.7)  \\   
               F_2(t) &= \textit{Exponential}(0.05) \\
               F_3(t) &= \textit{Weibull}(2.4, 11.7)  \\   
               F_4(t) &= \textit{Lognormal}(3, 1.6)  \end{aligned}$}& \raisebox{-.5\height}{
      \includegraphics[width=2.7in]{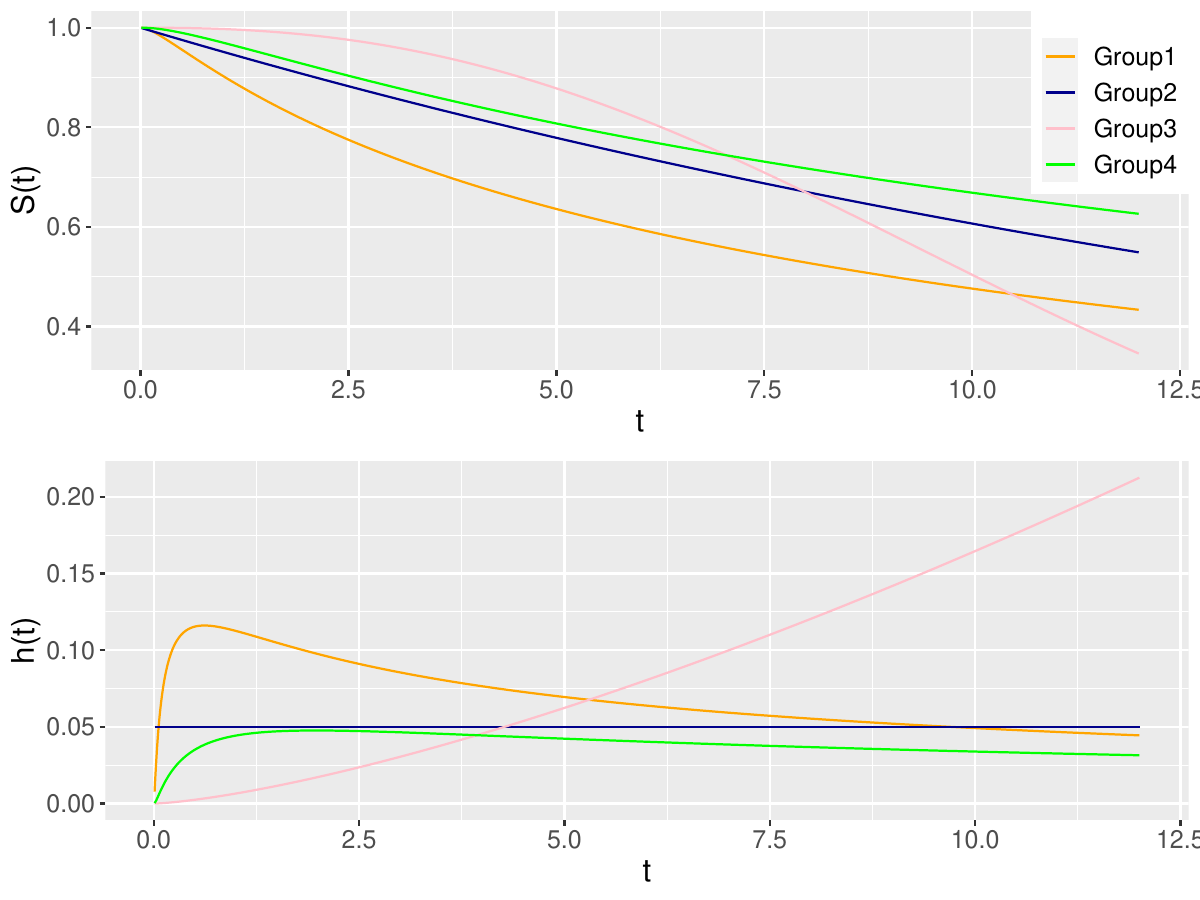}} \\
      \hline
    \end{tabular}
    \label{tab:Surv}
\end{table}

\subsection*{Simulation results under the null hypothesis}

\begin{figure}[H]
    \centering
    \includegraphics[width=\linewidth]{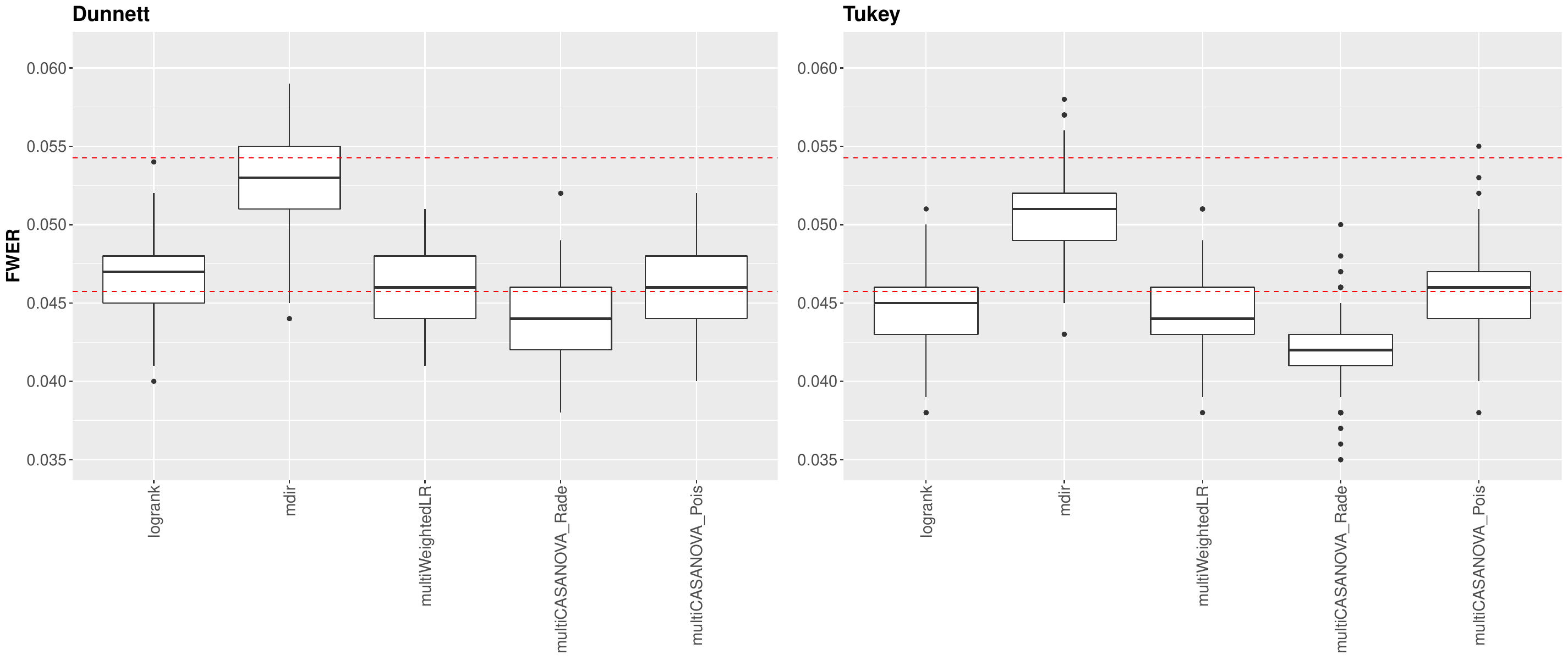}
    \caption{FWER under $\mathcal H_0$ for all settings for the Dunnett-type (left) and Tukey-type (right) contrast matrices. The dashed lines represent the borders of the binomial confidence interval $[4.57\%, 5.43\%]$}
    \label{fig:Null}
\end{figure}

Figure \ref{fig:Null} illustrates the familywise error rate (FWER) for all survival scenarios for the different contrast matrix types. We set the $\alpha$-level to $5\%$. The dashed lines represent the binomial confidence interval $[4.57\%, 5.43\%]$. 

For both contrast matrices, almost all methods control the FWER well. The adjusted mdir is the only test that is a little liberal when comparing the median to the global $\alpha$-level of $5\%$ for the Dunnett-type matrix. The new multiple-testing approaches are more conservative than the adjusted approaches, especially for the Tukey-type contrast matrices, with the multiWeightedLR being the most conservative.

\subsection*{Simulation results under the alternative hypothesis}
\begin{figure}[H]
    \centering
    \includegraphics[width=\linewidth]{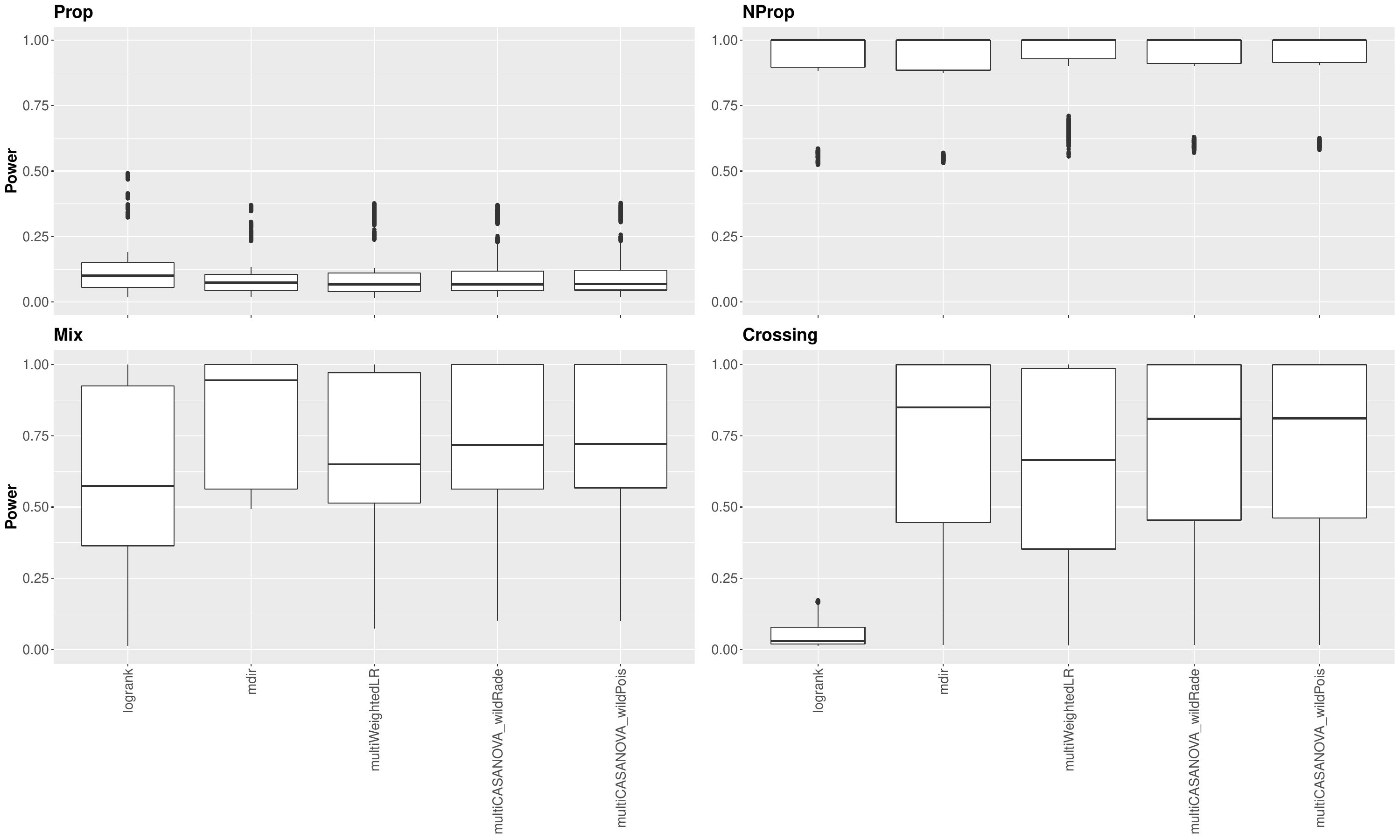}
    \caption{Local power over all tests under the alternative for Dunnett-type contrasts for all four scenarios (each boxplot contains 1136 data points).}
    \label{fig:Dun}
\end{figure}

\begin{figure}[H]
    \centering
    \includegraphics[width=\linewidth]{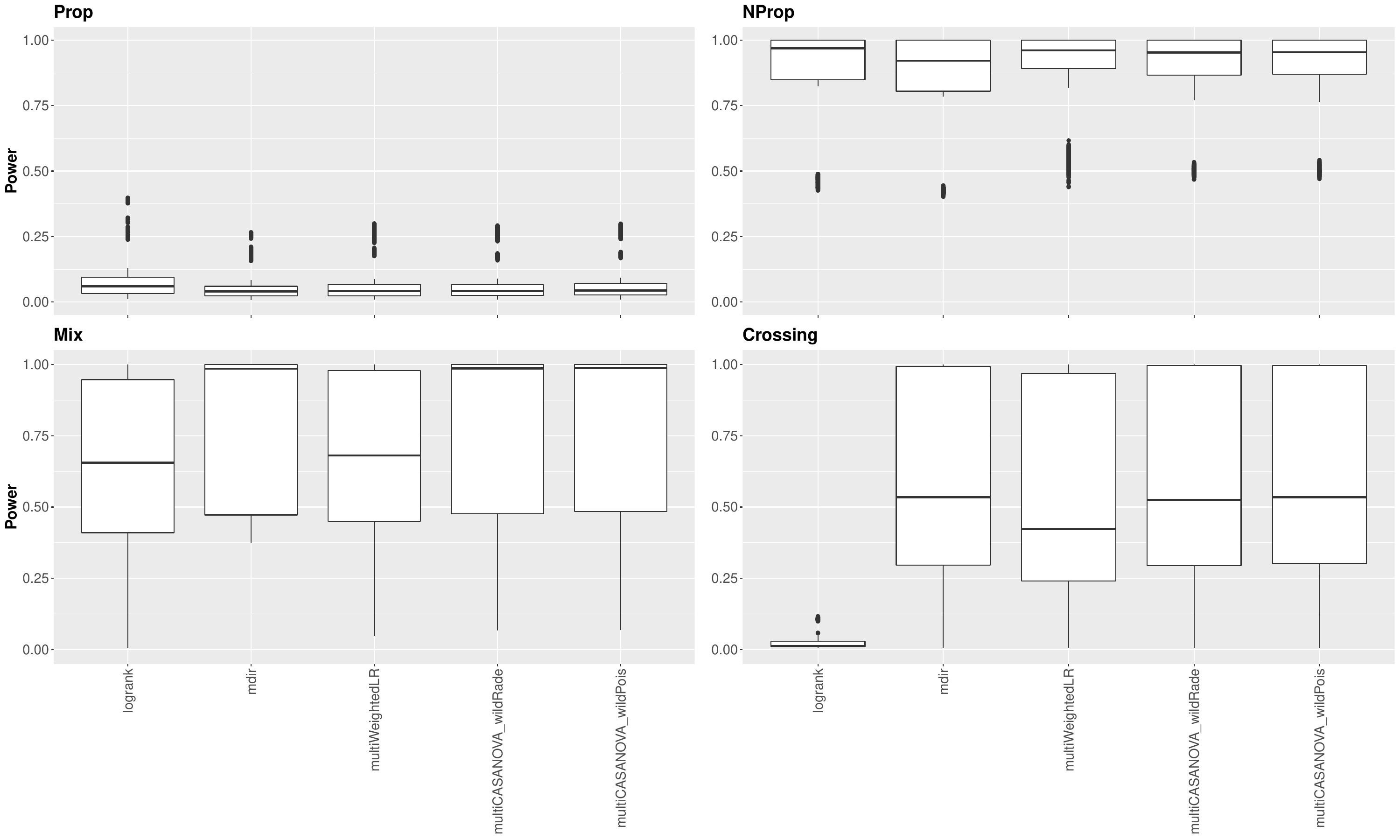}
    \caption{Local power over all tests under the alternative for Tukey-type contrasts for all four scenarios (each boxplot contains 2016 data points).}
    \label{fig:Tuk}
\end{figure}

We focused on the local decisions under the alternative hypothesis to assess the power. Figures \ref{fig:Dun} and \ref{fig:Tuk} illustrate the rejection rates when different survival distributions are present. Each figure consists of four subfigures, one for each scenario. It should be noted that a higher number of tests decreases the power of each local hypothesis. This property is visible in the plots showing generally higher power for the Dunnett plots than the Tukey plots. Besides that, the tests behave similarly for both contrast matrix types. The adjusted log-rank test is the most powerful in the setting with proportional hazards, while the other tests perform equally well. Under non-proportional but non-crossing hazards, all tests have a high power, with the new approaches yielding a slightly lower variability. In the crossing scenario, the log-rank test loses power drastically due to violating the PH assumption. The four approaches designed for nPH data have high power, with the multiWeightedLR being slightly less powerful than the other three tests. In the mixed setting, the adjusted mdir performs best in terms of power, followed by the three methods introduced in this paper. The log-rank test has the highest variability and the lowest median power.  

The rejection rates for the local tests with no difference in survival are depicted in Figures \ref{fig:DunFALSE} and \ref{fig:TukFALSE} in the Supplemental Material. Overall, the rejection rates among the approaches are similar, with lower rejection rates for the Tukey-type contrast matrices. Additionally, the power for each local test is provided in the tables in the Supplemental Material.

The adjusted mdir test performs best for two of the four settings considered. Considering that it showed slightly liberal behavior under the null hypothesis, these results should be interpreted carefully. The methods introduced in this publication yield robust results regarding power among the different scenarios. The adjusted log-rank test loses power dramatically in the scenario with crossing hazards.  

In the Appendix in Figure \ref{fig:Null50}-\ref{fig:Tuk50}, we present an additional analysis of the behavior of the different tests concerning the FWER and power for smaller sample sizes ($n=50$). The results indicate that the multiWeightedLR approach exhibits an inflated FWER, likely due to the normal approximation. In contrast, both multiCASANOVA bootstrap approaches maintain strong control over the family-wise error rate and consistently deliver good results in terms of power. The adjusted LR and mdir test still control the FWER but show increased variability in terms of power.

\section{Illustrative Data Example}

\begin{figure}[H]
    \centering
    \includegraphics[width=\linewidth]{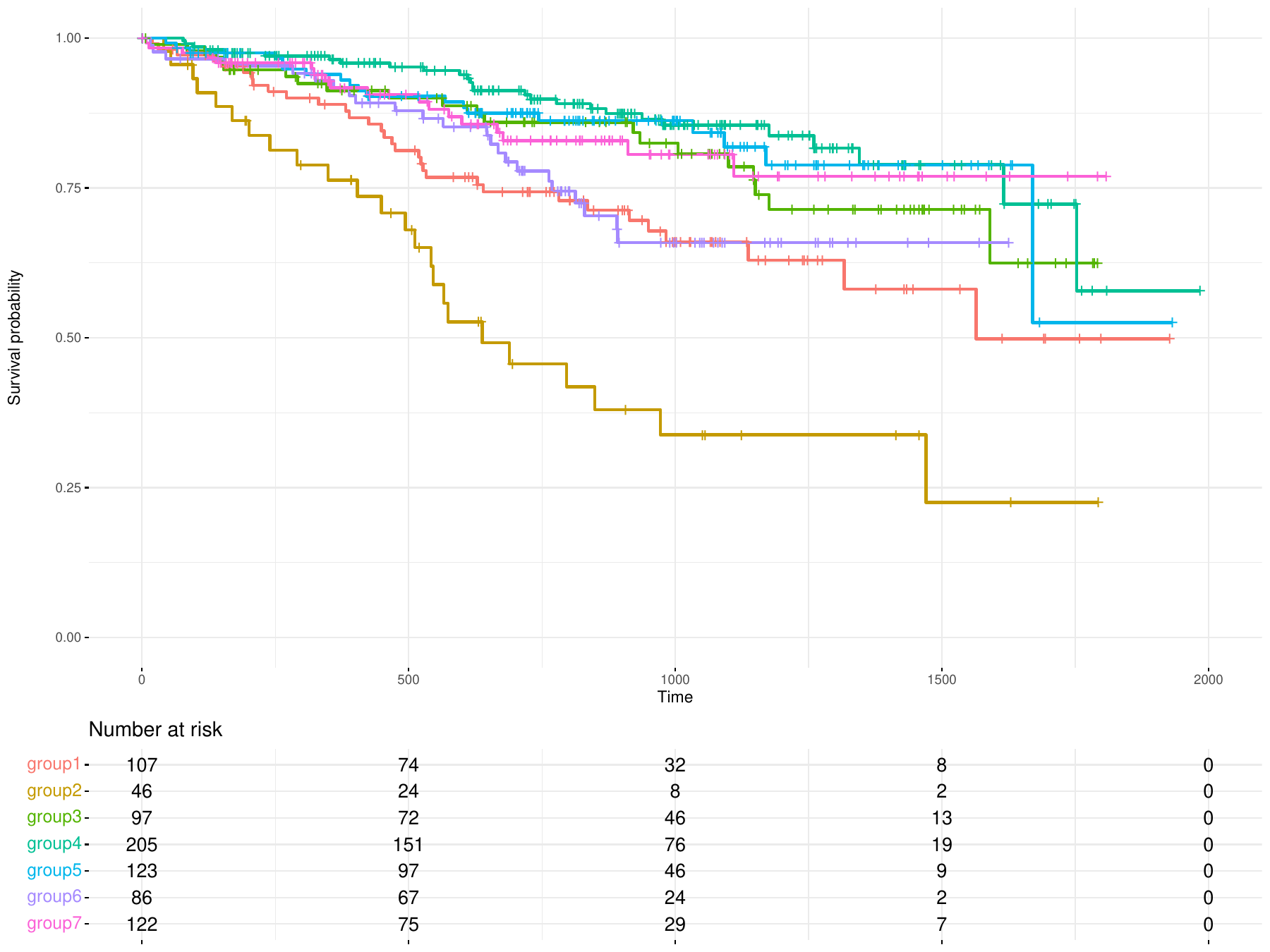}
    \caption{Kaplan-Meier plot of the seven treatment groups of patients with multiple myeloma (MM).}
    \label{fig:ExampleSurv}
\end{figure}

To illustrate the novel approaches on real-world data, we used publicly available data from the CoMMpass study (dbGaP accession: phs000748.v4.p3). This study is designed to associate clinical outcome with genetic profiles and contains longitudinal clinical and molecular data from multiple myeloma (MM) patients. Based on the transcriptional profile and the expression level of biologically relevant core machinery that plays a vital role in the stress response \cite{heynen2023sumoylation}, we clustered MM patients into seven groups. Figure \ref{fig:ExampleSurv} shows the Kaplan-Meier curves of the seven different groups. We assume that we are interested in comparing every group with one another and consider a Tukey design. The significance level was set to $\alpha = 0.05$ with a total of $21$ tests. The corrected significance level is thus $\alpha_{\text{Bonferroni}} = 0.0024$.

 By examining the survival curves, we anticipate that the methods will identify a significant overall difference among the groups. Specifically, we expect group 2 to differ from the other groups. However, we do not expect to see any differences among groups 3, 4, and 5. We applied all testing procedures described in this paper to investigate these premises. For all approaches combining multiple weighted log-rank tests, we included the $w^{(1)} = 0$ and $w^{(2)} = 1 - 2\widehat{S}(t_i)$. We set the number of resampling iterations to $1000$ for all resampling-based approaches.

The detailed results are listed in the Supplemental Material Table \ref{tab:pvalsExample}. The adjusted log-rank and mdir test detected six significant differences between groups, while the three new methods only detected five (see Table \ref{tab:nbrReject}). The found differences are consistent among the methods. All tests found the pair-wise differences between groups two and three, four, five, and seven, as well as between groups one and four, to be significant. A significant result for the comparison between groups two and six was only found by the adjusted LR test and the adjusted mdir test.

\begin{table}[H]
\footnotesize
\centering
\setlength{\tabcolsep}{5pt} 
\renewcommand{\arraystretch}{1.2} 
\caption{Number of rejected local null hypotheses for each of the tests}
\label{tab:nbrReject}

\begin{tabular}{lccccc}
  \toprule
  & \textbf{logrank} & \textbf{mdir} & \textbf{multiWeightedLR} & \textbf{multiCASANOVA\_wildRade} & \textbf{multiCASANOVA\_wildPois} \\ 
  \midrule
  \textbf{Nbr. of rejected} $H_0^{j_1j_2}$ & 6 & 6 & 6 & 5 & 5 \\ 
  \bottomrule
\end{tabular}
\end{table}

All tests could reject the global hypothesis of any difference between groups as well. 
In summary, we could show that in the case of a real-world application, the results are consistent with the results of the adjusted log-rank test. 

\section{Discussion}
We explored various statistical methods for addressing multiple contrast problems with time-to-event endpoints, including traditional and newly developed approaches. To assess the approaches' performance, we compared the Family-Wise Error Rate (FWER) control and the power performance of these methods under different survival scenarios. The results of our simulation study and real-world data application provide valuable insights into the strengths and limitations of each approach. 

Most methods maintain adequate control of the FWER. The adjusted mdir test exhibited a slightly liberal behavior, particularly for Dunnett-type contrasts. This deviation suggests that while the adjusted mdir test might be powerful, it occasionally exceeds the acceptable error rate, which warrants caution in its interpretation under null conditions. On the other hand, the multiWeightedLR and multiCASANOVA methods were generally more conservative, particularly for Tukey-type contrast matrices. This conservativeness could imply a lower risk of Type I errors but may come at the cost of reduced statistical power.

Under alternative hypotheses, the power analysis revealed notable differences in the tests' performance depending on the survival scenario. For proportional hazards, the adjusted log-rank test showed the highest power, outperforming the other methods. Under non-proportional and non-crossing hazards, we could observe high power among all tests, with the new approaches showing slightly lower variability. The robustness of these methods suggests that they are suitable choices when proportional hazards are not guaranteed. The log-rank test's power decreased drastically in the specific case of crossing hazards.
In contrast, the four approaches specifically designed for non-proportional hazards (adjusted mdir, multiWeightedLR, and the two multiCASANOVA variants) maintained high power, confirming their utility in these settings. Finally, the adjusted mdir test outperformed other methods in the mixed scenario, achieving the best power performance. Although slightly less powerful, the new methods provided more consistent results across different scenarios, highlighting their robustness.

The results suggest potential areas for further methodological improvements. While the Bonferroni correction is widely used for controlling type I error rates, its conservative nature may result in lower power, particularly in settings with many comparisons. More sophisticated adjustment techniques, like the Holm procedure, could better balance error rate control and power, as discussed in previous studies.

Additionally, evaluating the performance of these methods in unbalanced designs could provide a more comprehensive understanding of their behavior in practical applications. This could be particularly interesting since \cite{munko2024} showed that such conditions could boost the power of specific local designs.

In the illustrative data example involving patients with multiple myeloma, the new methods produced consistent results with those obtained from the adjusted log-rank tests. Although the novel approaches identified less significant differences than the traditional methods, their findings were largely aligned, underscoring their reliability in practical scenarios. This consistency and the conservative behavior in terms of FWER control suggest that the new methods still offer a robust alternative for analyzing time-to-event data in clinical studies. Future research would include more efficient exploitation of the FWER for the new approaches, e.g., by incorporating closed testing approaches. In general, it is essential to critically assess whether a higher number of statistically significant results truly reflects a superior testing approach, as statistical significance does not inherently equate to clinical relevance.

\section*{Acknowledgements}
This work has been partly supported by the Research Center Trustworthy Data Science and Security (https://rc-trust.ai), one of the Research Alliance centers within https://uaruhr.de. The authors gratefully acknowledge the computing time provided on the Linux HPC cluster at Technical University Dortmund (LiDO3), partially funded in the course of the Large-Scale Equipment Initiative by the German Research Foundation (DFG) as project 271512359. Moreover, the work of Marc Ditzhaus and  Markus Pauly was supported by the joined DFG Sachbeihilfe-project (project number 352692197). These data were generated as part of the Multiple Myeloma Research Foundation CoMMpass [SM] (Relating Clinical Outcomes in MM to Personal Assessment of Genetic Profile) study (www.themmrf.org).

In memory of our esteemed colleague, Marc Ditzhaus, who passed away in September 2024. His invaluable input will always be remembered, and he will be deeply missed. 

\bibliographystyle{unsrt}  
\bibliography{references}  

\newpage

\section*{Supporting Information}

\begin{figure}[H]
    \centering
    \includegraphics[width=\linewidth]{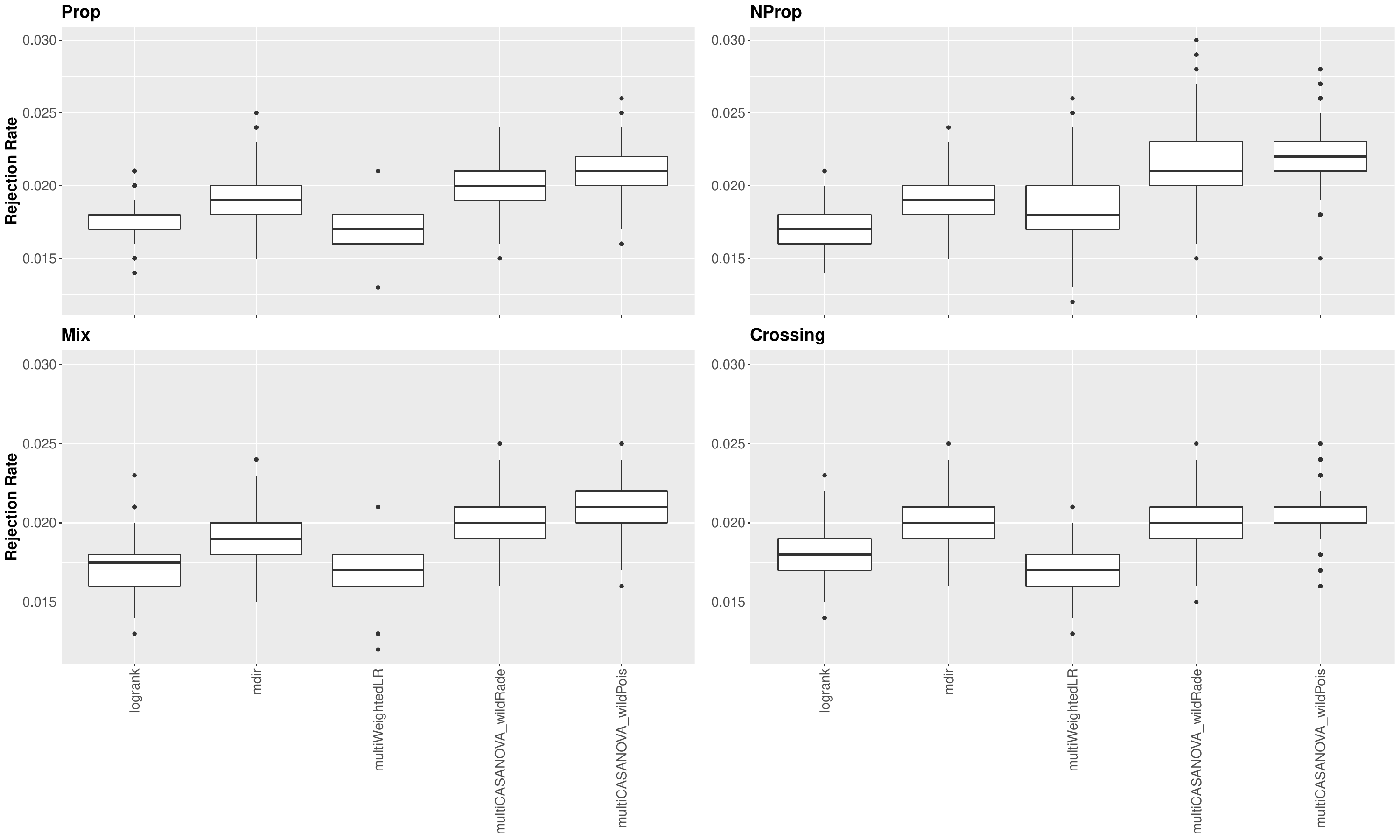}
    \caption{Rejection rate of the local tests with no difference in survival for the Dunnett-type matrix.}
    \label{fig:DunFALSE}
\end{figure}

\begin{figure}[H]
    \centering
    \includegraphics[width=\linewidth]{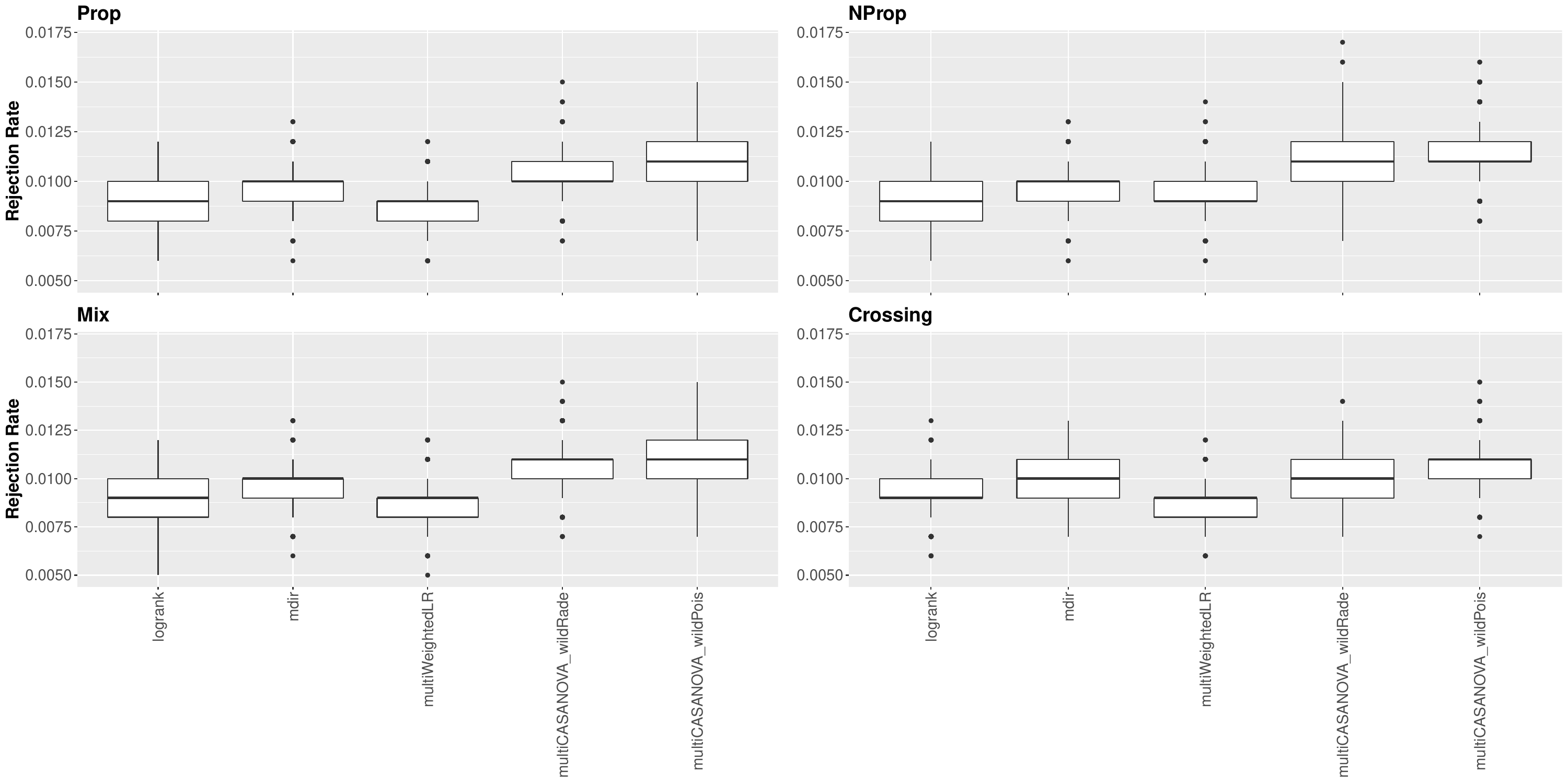}
    \caption{Rejection rate of the local tests with no difference in survival for the Tukey-type matrix.}
    \label{fig:TukFALSE}
\end{figure}

\begin{table}[ht]
\centering
\caption{p-values for the local comparisons in the data example (Section 4). Significant tests to level $\alpha$ or $\alpha_{\text{Bonferroni}}$ are highlighted in bold font.}
\begin{tabular}{rrrrrr}
  \hline
 & log-rank & mdir & multiWeightedLR & multiCASANOVA\_wildRade & multiCASANOVA\_wildPois \\ 
  \hline
2 - 1 & 0.196 & 0.411 & 0.130 & 0.301 & 0.304 \\ 
  3 - 1 & 0.100 & 0.152 & 0.415 & 0.699 & 0.727 \\ 
  4 - 1 & \textbf{0.001} & \textbf{0.003} & \textbf{0.013} & \textbf{0.030} & \textbf{0.023} \\ 
  5 - 1 & \textbf{0.010} & \textbf{0.028} & 0.135 & 0.203 & 0.210 \\ 
  6 - 1 & 0.502 & 0.515 & 0.981 & 1.000 & 1.000 \\ 
  7 - 1 & 0.040 & 0.136 & 0.366 & 0.571 & 0.566 \\ 
  3 - 2 & \textbf{<0.001} & \textbf{0.001} & \textbf{0.006} & \textbf{0.012} & \textbf{0.005} \\ 
  4 - 2 & \textbf{<0.001} & \textbf{<0.001} & \textbf{0.001} & \textbf{0.003} & \textbf{<0.001} \\ 
  5 - 2 & \textbf{<0.001} & \textbf{<0.001} & \textbf{0.003} & \textbf{0.007} & \textbf{0.001} \\ 
  6 - 2 & \textbf{0.002} & \textbf{0.001} & \textbf{0.040} & 0.092 & \textbf{0.091} \\ 
  7 - 2 & \textbf{<0.001} & \textbf{<0.001} & \textbf{0.006} & \textbf{0.011} & \textbf{0.003} \\ 
  4 - 3 & 0.406 & 0.217 & 0.821 & 0.980 & 0.991 \\ 
  5 - 3 & 0.612 & 0.759 & 0.999 & 1.000 & 1.000 \\ 
  6 - 3 & 0.242 & 0.534 & 0.929 & 0.985 & 0.995 \\ 
  7 - 3 & 0.889 & 0.943 & 1.000 & 1.000 & 1.000 \\ 
  5 - 4 & 0.885 & 0.990 & 0.956 & 0.981 & 0.993 \\ 
  6 - 4 & 0.034 & 0.046 & 0.175 & 0.415 & 0.414 \\ 
  7 - 4 & 0.777 & 0.768 & 0.760 & 0.922 & 0.942 \\ 
  6 - 5 & 0.195 & 0.329 & 0.649 & 0.776 & 0.789 \\ 
  7 - 5 & 0.959 & 1.000 & 0.999 & 1.000 & 1.000 \\ 
  7 - 6 & 0.424 & 0.248 & 0.907 & 0.941 & 0.959 \\ 
   \hline
\end{tabular}
\label{tab:pvalsExample}
\end{table}

\begin{figure}[H]
    \centering
    \includegraphics[width=\linewidth]{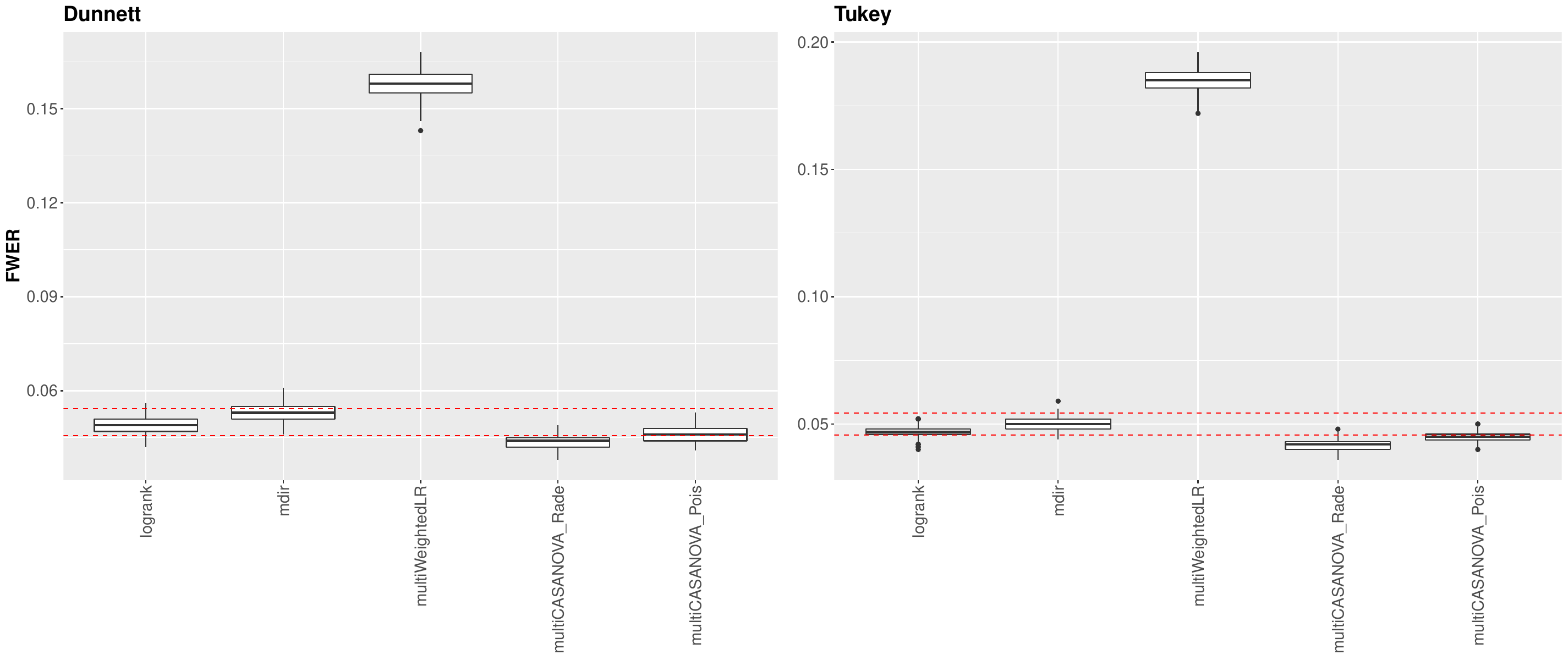}
    \caption{FWER under $\mathcal H_0$ for all settings for the Dunnett-type (left) and Tukey-type (right) contrast matrices for $n=50$. The dashed lines represent the borders of the binomial confidence interval $[4.57\%, 5.43\%]$}
    \label{fig:Null50}
\end{figure}

\begin{figure}[H]
    \centering
    \includegraphics[width=\linewidth]{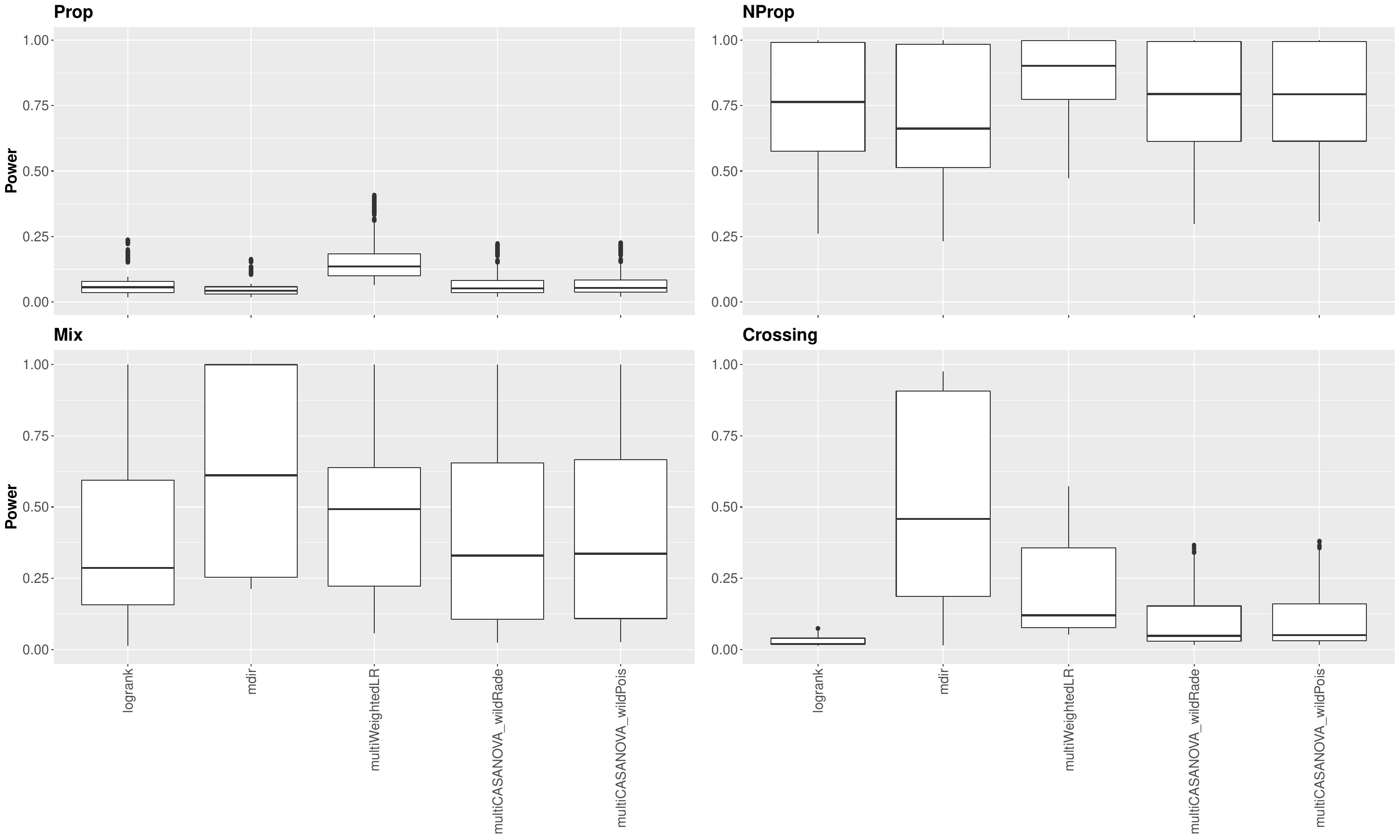}
    \caption{Local power over all tests under the alternative for $n=50$ for Dunnett-type contrasts for all four scenarios (each boxplot contains 1136 data points).}
    \label{fig:Dun50}
\end{figure}

\begin{figure}[H]
    \centering
    \includegraphics[width=\linewidth]{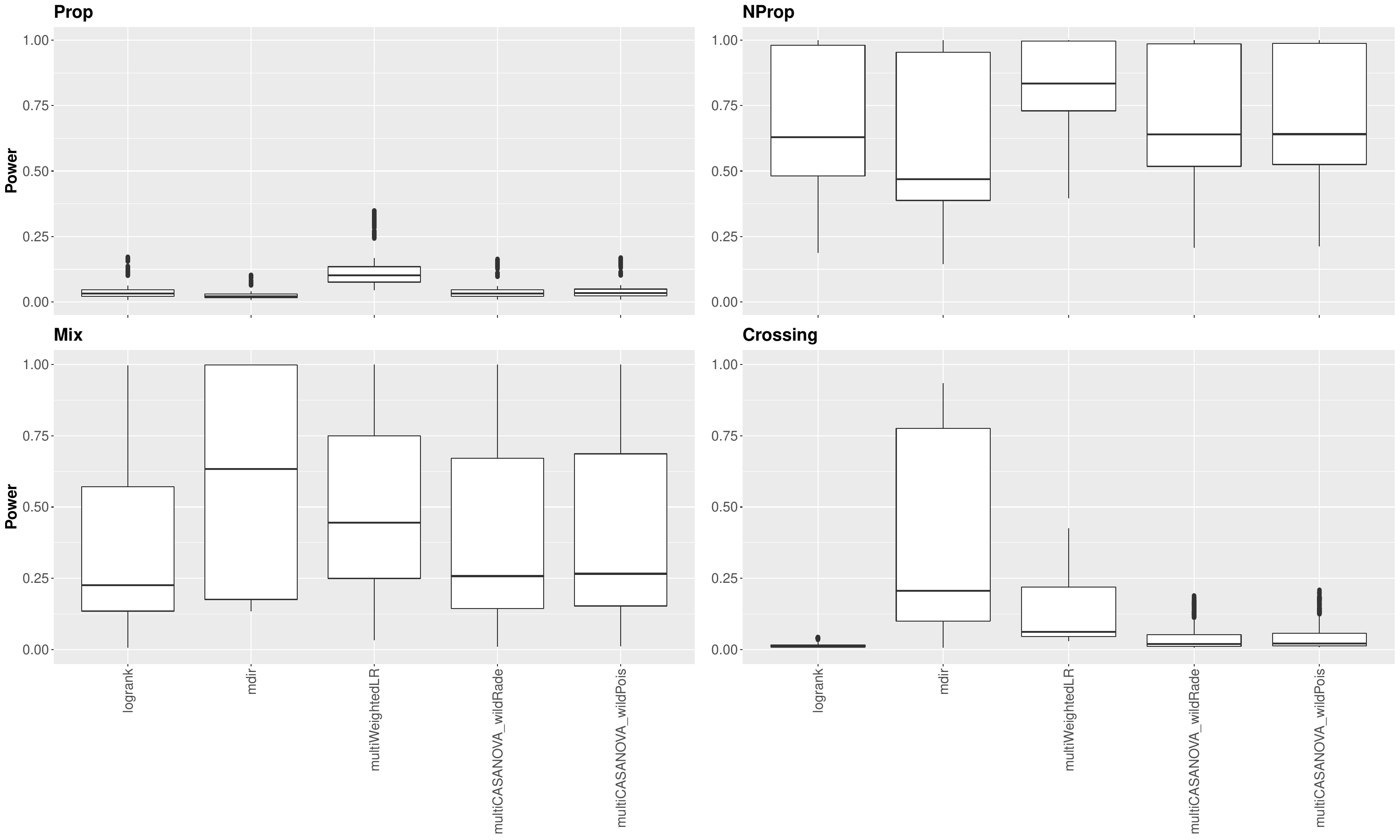}
    \caption{Local power over all tests under the alternative for $n=50$ for Tukey-type contrasts for all four scenarios (each boxplot contains 2016 data points).}
    \label{fig:Tuk50}
\end{figure}

\end{document}